\documentclass[nonacm, sigconf]{acmart}

\usepackage[english]{babel}
\usepackage[utf8]{inputenc}
\usepackage{tikz-cd}
\usepackage{lipsum}
\usepackage{wasysym}
\usepackage{multirow}
\usepackage{tcolorbox}

\usepackage{hyperref}
\usepackage{array}                  
\usepackage{verbatim}
\newcolumntype{L}[1]{>{\raggedright\let\newline\\\arraybackslash\hspace{0pt}}m{#1}}
\newcolumntype{C}[1]{>{\centering\let\newline\\\arraybackslash\hspace{0pt}}m{#1}}
\newcolumntype{R}[1]{>{\raggedleft\let\newline\\\arraybackslash\hspace{0pt}}m{#1}}
\newcolumntype{J}[1]{>{\let\newline\\\arraybackslash\hspace{0pt}}m{#1}}

\AtBeginDocument{%
  \providecommand\BibTeX{{%
    \normalfont B\kern-0.5em{\scshape i\kern-0.25em b}\kern-0.8em\TeX}}}

\begin{document}



\title{Event-Driven Inconsistency Detection Between UML Class and Sequence Diagrams}

\author{Luan Lazzari}
\email{luanlazzari@edu.unisinos.br}
\affiliation{%
  \institution{Universidade do Vale do Rio dos Sinos}
  \city{São Leopoldo}
  \state{Rio Grande do Sul}
  \country{Brazil}
}

\author{Kleinner Farias}
\email{kleinnerfarias@unisinos.br}
\affiliation{%
  \institution{Universidade do Vale do Rio dos Sinos}
  \city{São Leopoldo}
  \state{Rio Grande do Sul}
  \country{Brazil}
}

\begin{abstract}
Modeling is a central and demanding activity in software engineering that requires skills such as abstraction, consistency maintenance, and precise communication. These skills are difficult to master and even harder to teach effectively. Educators and students often struggle to understand and manage inconsistencies that arise during the modeling process. To address this challenge, we present \texttt{Harmony Validator}, a tool integrated as a plugin for the Papyrus modeling environment, designed to automatically detect and report inconsistencies in UML models, including class and sequence diagrams. The tool adopts an event-driven architecture that continuously monitors modeling actions and notifies users of emerging inconsistencies in real time. This approach enhances awareness of model integrity and supports the iterative refinement of design artifacts. The paper describes the architecture, detection mechanisms, and usage scenarios of Harmony Validator. It also includes a case study conducted with students in a software engineering course to evaluate the perceived usefulness and benefits of UML modeling in teaching and learning. Our results indicate that Harmony Validator fosters a better understanding of model consistency and promotes reflective learning practices in software modeling education.
\end{abstract}


\maketitle

\section{Introduction}
\label{sec:intro}

High-quality documentation significantly increases the success of software projects~\cite{dzidek2008realistic, lethbridge2003software, tryggeseth1997report, romeo2025uml}. This is particularly evident for visual documentation based on the Unified Modeling Language (UML) \cite{specificationuml}, which remains a central artifact in both academia and industry \cite{junior2021survey,junior2022use}. UML diagrams foster shared understanding, encourage design discussions, and support reasoning about system structure and behavior~\cite{romeo2025uml,gonccales2019comparison}. Recent evidence shows that UML is regaining attention as a standard medium for documenting and communicating software design decisions~\cite{romeo2025uml}. In educational settings, UML enhances the teaching of abstraction, decomposition, and communication skills. 
In industry, consistent UML documentation \cite{farias2016empirical,weber2016detecting,farias2012evaluating} improves knowledge transfer across teams and contributes to software maintainability and quality~\cite{jolak2022influence, arisholm2006impact}. Empirical findings demonstrate that developers achieve higher functional correctness and introduce fewer defects during maintenance when UML models are up-to-date and consistent~\cite{dzidek2008realistic, arisholm2006impact,gonccales2019comparison,oliveira2009flexible}. 
These studies reinforce the importance of \textit{inconsistency detection} to sustain the reliability of UML models and to preserve their benefits for software comprehension and evolution.

The \texttt{Harmony Validator} tool (Table \ref{tab:metadata}) was conceived to address an ongoing challenge in model-driven engineering, i.e., the \textit{automatic detection of inconsistencies across UML diagrams} \cite{torre2020uml,torre2018systematic}. As software systems evolve, disagreements often arise between structural (class) and behavioral (sequence) diagrams, leading to conceptual mismatches that compromise model integrity. 
Manual validation of these relationships is error-prone and time-consuming, particularly in collaborative modeling environments. 
Existing tools typically detect inconsistencies through post-hoc or batch verification, providing limited real-time support \cite{marchezan2024tool,wu2017maxuse}. 
\texttt{Harmony Validator} fills this gap through an \textit{event-driven architecture} that continuously monitors modeling actions and reports inconsistencies as they occur. 
This design supports both students and practitioners in identifying, interpreting, and resolving inconsistencies early in the modeling process, reducing cognitive load \cite{bolzan2025investigating,segalotto2023effects} and enhancing design correctness.

\begin{table*}[!ht]
\centering
\begin{tabular}{|p{6.5cm}|p{8.5cm}|}
\hline
\textbf{Code metadata item} & \textbf{Description} \\
\hline
Current code version & V1.0 \\
\hline
Permanent link to code/repository used for this code & 
\url{https://github.com/luanlazz/uml-harmony-validator-service} \\
version & \url{https://github.com/luanlazz/uml-harmony-validator-plugin}\\
\hline
Legal Code License   & MIT License \\
\hline
Code versioning system used & git \\
\hline
Software code languages, tools, and services used & Java, Kafka, Redis \\
\hline
Compilation requirements, operating environments and dependencies &  Docker, Java 17, Apache Maven 3.5+ \\
\hline
Link to developer documentation/manual & 
\url{https://github.com/luanlazz/uml-harmony-validator-service/blob/master/readme.md}
\url{https://github.com/luanlazz/uml-harmony-validator-plugin/blob/main/README.md}\\
\hline
Support email for questions & luanlazzari@gmail.com \\
\hline
\end{tabular}
\caption{Description of the source code metadata.}
\label{tab:metadata}
\end{table*}

Users employ the \texttt{Harmony Validator} within the Papyrus\footnote{Papyrus: https://eclipse.dev/papyrus/index.html} modeling environment to create and refine UML class and sequence diagrams while the tool continuously monitors model changes and automatically detects inconsistencies such as undefined operations, broken associations, or mismatched messages. Real-time notifications highlight the affected elements, prompting immediate user attention and fostering iterative correction and design coherence. In educational settings, \texttt{Harmony Validator} has been used in laboratory sessions to train students in identifying and resolving inconsistencies, improving modeling fluency and awareness of inter-diagram dependencies. Built on robust foundations in model validation and event-driven systems, the tool extends Papyrus—based on the Eclipse Modeling Framework\footnote{Eclipse Modeling Framework: https://eclipse.dev/emf/} (EMF) and integrates Apache Kafka\footnote{Apache Kafka: https://kafka.apache.org/} for efficient event handling. Implemented in Java language and deployed via Docker containers, \texttt{Harmony Validator} ensures interoperability, reproducibility, and scalability, enabling its seamless integration into both academic experiments and industrial modeling workflows.

The \texttt{Harmony Validator} distinguishes itself from existing approaches \cite{sultan2024ai,marchezan2024exploring,marchezan2024tool} in three main ways: (i) It introduces an \textit{event-driven, continuous-monitoring paradigm}, enabling real-time inconsistency detection rather than relying on static, batch-oriented validation; (ii) It performs \textit{bi-directional consistency checking} between structural and behavioral UML diagrams, ensuring semantic alignment between class definitions and interaction flows; and (iii) It was designed as both a \textit{pedagogical and research-oriented tool}, supporting empirical investigations into modeling practices while also assisting practitioners in maintaining model integrity.

This study is structured as follows: Section~\ref{sec:software-description} describe the proposed tool, outlining its architecture, functionalities and practical use case. Section \ref{sec:illustrative-examples} presents an practical usage example of the proposed tool. Section \ref{sec:impact} outlines the impact of the proposed tool. Finally, Section \ref{sec:conclusions} introduces some concluding remarks and future plans.

\section{Software description}
\label{sec:software-description}

\subsection{Software architecture}

\begin{figure*}[!ht]
    \centering
    \includegraphics[width=0.95\textwidth]{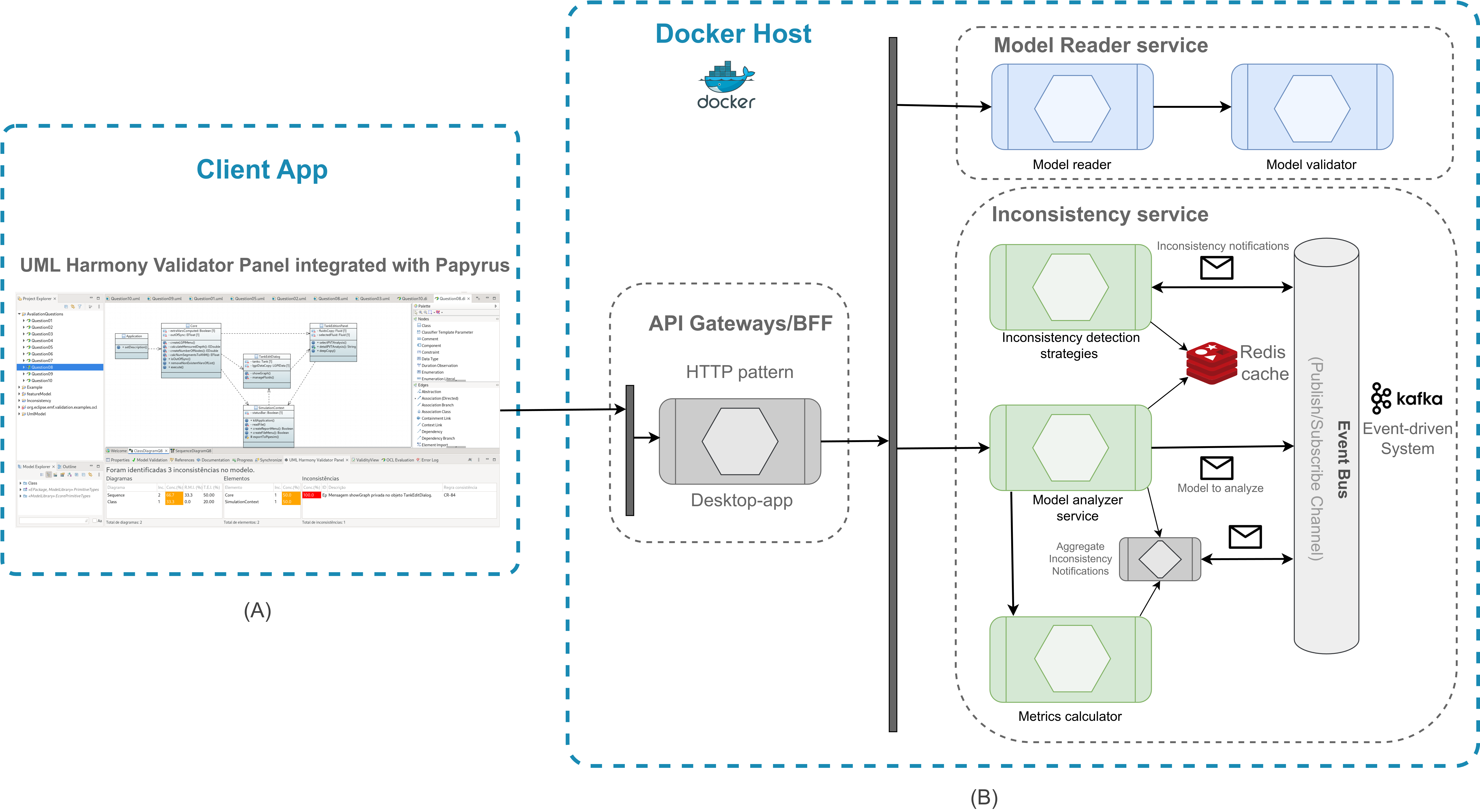}
    \caption{Event-driven architecture of the Harmony Validator. 
    (A) Client App integrated with Papyrus modeling environment. 
    (B) Docker Host containing model reader, inconsistency detection, and event bus services.}
    \label{fig:architecture}
\end{figure*}

Figure~\ref{fig:architecture} presents an overview of the \texttt{Harmony Validator} architecture, which follows an \textit{event-driven architectural style} \cite{cabane2024impact,lazzari2023uncovering,lazzari2023event}. The system is divided into two main modules: the \textit{Client App} and the \textit{Docker Host} environment. This separation ensures modularity, scalability, and ease of deployment across different platforms.

In Figure~\ref{fig:architecture}(A), the \textbf{Client App} corresponds to a plugin integrated into the Papyrus modeling environment, where users perform UML modeling activities. Within Papyrus, the \texttt{Harmony Validator} adds a custom panel that continuously monitors and reports inconsistencies detected in the UML model. 
The \texttt{Harmony Validator} interface presents feedback to the user through visual indicators that identify affected elements and diagrams. 
This integration enables users to maintain focus on modeling tasks while receiving real-time feedback on inconsistencies.

In Figure~\ref{fig:architecture}(B), the \textbf{Docker Host} contains a set of services deployed as independent containers, connected through an asynchronous \textbf{Event Bus}. 
The event-driven layer is responsible for receiving and processing model modification events. 
The \textit{Model Reader Service} retrieves the updated model data from Papyrus and publishes the modification events to the system. 
The \textit{Inconsistency Service} subscribes to these events, analyzes the model state, and identifies inconsistencies across UML diagrams. 
Detected inconsistencies are then cached using \textbf{Redis} for quick retrieval and aggregated notifications. 
All communication between the client and backend follows an HTTP pattern implemented through an \textbf{API Gateway/BFF (Backend for Frontend)} module, which mediates requests from the desktop client to the microservices running in Docker. 
This architecture promotes decoupling between components, supports parallel modeling scenarios, and allows for seamless scalability.

\subsection{Software functionalities}

The \texttt{Harmony Validator} provides a set of major functionalities designed to support inconsistency detection and enhance user awareness during modeling activities:

\begin{enumerate}
    \item \textit{Real-time inconsistency detection:} Continuously monitors user actions in Papyrus and detects inconsistencies between UML class and sequence diagrams as they occur.
    \item \textit{Event-driven notification system:} Uses an asynchronous event bus to deliver immediate alerts through the modeling interface when inconsistencies are found.
    \item \textit{Distributed microservice architecture:} Each component runs independently in Docker containers, ensuring scalability and modular deployment.
    \item \textit{Data caching and aggregation:} Employs Redis for caching detected inconsistencies and aggregating results, reducing computation time for recurring analyses.
    \item \textit{Seamless integration with Papyrus:} Provides an intuitive panel that allows users to inspect inconsistency details and navigate to affected elements within the modeling environment.
\end{enumerate}

These functionalities collectively enable developers and learners to maintain high levels of model consistency while improving comprehension of inter-diagram dependencies.

\subsection{Sample use case}

A typical use case involves a user editing a UML class and sequence diagram in Papyrus. 
As the user introduces changes---for instance, renaming an operation or modifying an association---the \texttt{Harmony Validator} automatically detects these modifications through the \textit{Model Reader Service}. 
Events are published to the \textit{Event Bus}, triggering the \textit{Inconsistency Service} to re-evaluate the affected diagrams. 
If a mismatch or undefined reference is identified, the system caches the result and notifies the user through the Papyrus panel. 
The user can then click the inconsistency message to navigate directly to the problematic element, correct the model, and immediately see the inconsistency resolved. 
This feedback loop exemplifies the tool’s ability to \textit{support reflective modeling practices} and to reinforce the understanding of consistency rules during both educational and industrial modeling activities.

\section{Illustrative Examples}
\label{sec:illustrative-examples}

\begin{figure*}[!ht]
    \centering
    \includegraphics[width=0.97\textwidth]{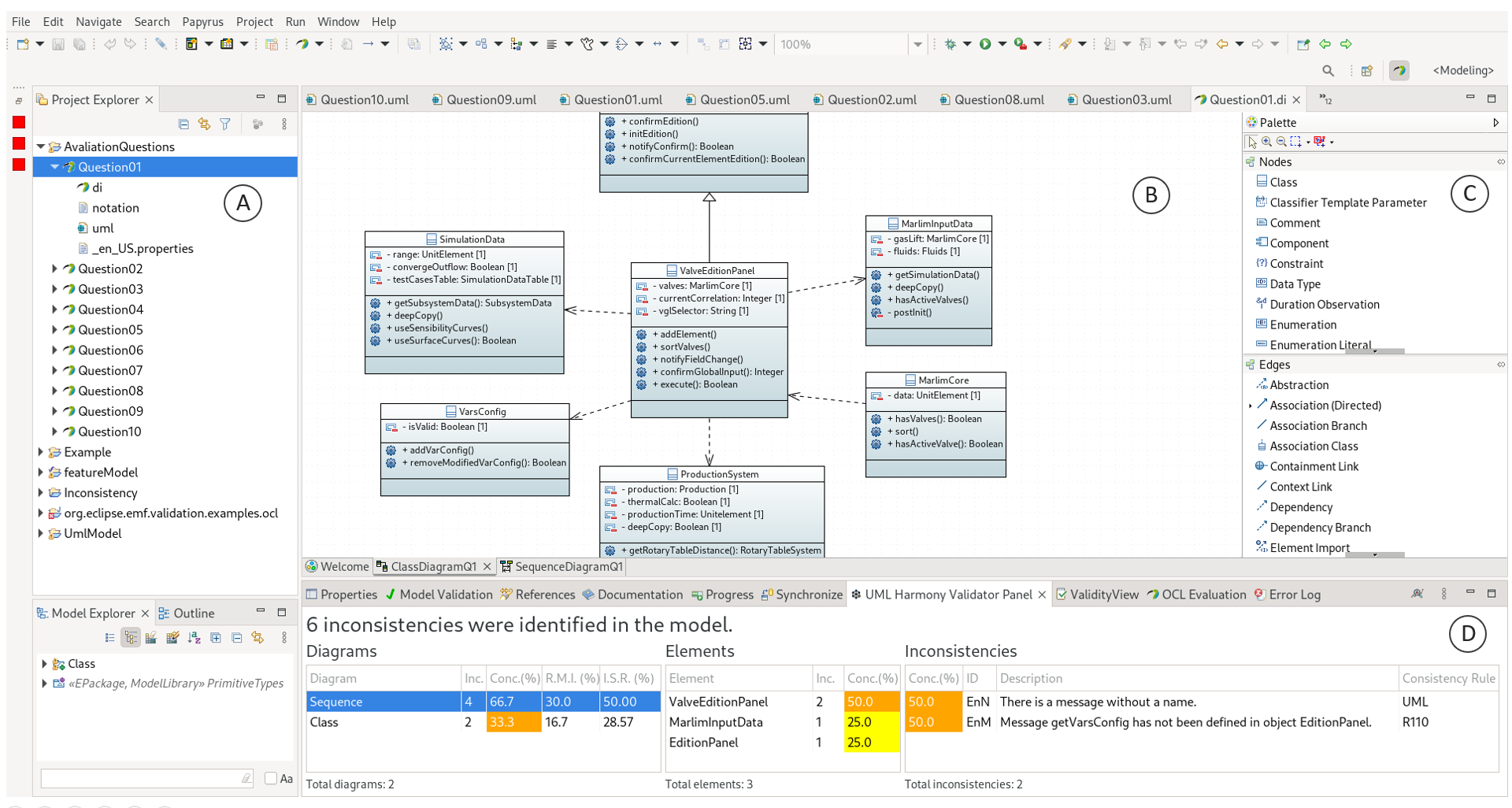}
    \caption{Illustrative example of the \texttt{Harmony Validator} integrated into Papyrus. 
    (A) Project Explorer for model organization. 
    (B) UML diagram editor showing model elements. 
    (C) Modeling palette with available UML elements. 
    (D) \texttt{Harmony Validator} panel displaying detected inconsistencies and corresponding rules.}
    \label{fig:interface}
\end{figure*}

Figure~\ref{fig:interface} illustrates the main functionalities of the \texttt{Harmony Validator} tool embedded in the \textbf{Papyrus} modeling environment. 
Part~(A) shows the \textbf{Project Explorer}, where users organize and access multiple UML projects and diagrams. 
In this example, the project \textit{AvaliationQuestions} contains a set of UML class and sequence diagrams used for teaching and experimentation. 
\texttt{Harmony Validator} operates in the background, continuously monitoring changes to these diagrams. 
Whenever a model element is added, modified, or removed, the system triggers an event that is analyzed by the event-driven architecture described previously. 
This real-time monitoring ensures that inconsistencies are promptly detected and reported without interrupting the modeling workflow.

Part~(B) depicts the \textbf{modeling canvas}, where the user edits the UML diagrams. 
In this workspace, developers can create and manipulate classes, associations, and behavioral interactions. 
The \texttt{Harmony Validator} automatically identifies inconsistencies that arise during these edits—for example, operations that are invoked in a sequence diagram but not defined in the corresponding class diagram. 
This continuous analysis exemplifies the tool’s primary functionality: maintaining semantic and structural consistency between UML views. 
The modeler can work naturally in the diagram editor, while the tool’s backend services detect and store data about inconsistencies, enabling transparent, proactive validation during design activities.

Part~(C) presents the \textbf{modeling palette}, which provides modeling elements such as \textit{Class}, \textit{Association}, and \textit{Enumeration}. 
As users drag and drop new elements into the diagram, the \texttt{Harmony Validator} automatically registers the modifications through the event bus, ensuring immediate propagation of changes to the inconsistency detection service. 
Finally, Part~(D) highlights the \textbf{Harmony Validator Panel}, where the results of the analysis are presented. 
The panel lists all detected inconsistencies, their type, location, and associated consistency rule. 
In the illustrated example, six inconsistencies were identified, including a message without a name and an undefined operation. 
Users can click on each entry to navigate directly to the affected element, view detailed descriptions, and correct the detected inconsistencies. 
This interactive feedback loop demonstrates how the \texttt{Harmony Validator} integrates real-time detection, visualization, and corrective guidance to promote high-quality UML modeling practices.

\section{Impact} 
\label{sec:impact}

The \texttt{Harmony Validator} tool creates new opportunities for research and practice in model-driven engineering. 
Empirical evidence suggests a renewed use of UML models in open-source projects hosted on platforms such as GitHub, where visual and textual representations coexist through tools like \textit{PlainUML}~\cite{romeo2025uml}. 
\texttt{Harmony Validator} supports the construction of consistent class and sequence diagrams, providing a foundation for automatically transforming them into textual UML representations (e.g., \textit{pUML}) using large language models (LLMs). 
This opens an innovative research direction for leveraging consistent visual models to automatically generate, refine, and synchronize textual and code-based representations. 
Such integration between consistency-validated UML diagrams and AI-driven code generation promotes more traceable, verifiable, and maintainable software artifacts, thereby extending the reach of empirical research on UML’s resurgence in open-source ecosystems.

From a research perspective, \texttt{Harmony Validator} advances experimental studies in software modeling by enabling fine-grained analysis of how inconsistencies emerge and are resolved during modeling activities. 
The tool’s event-driven logging infrastructure records real-time user interactions, allowing researchers to explore new questions about cognitive load \cite{bolzan2025investigating, gonccales2021measuring}, decision-making, and the effectiveness of automated feedback in maintaining model quality. 
These data can be used to design and reproduce experiments on how users respond to different types of inconsistencies or how automated validation influences modeling performance. 
Moreover, the combination of \texttt{Harmony Validator}’s architecture and its potential for data analytics paves the way for adaptive modeling environments where intelligent assistants or LLMs reason over event streams to suggest corrective actions, anticipate conflicts, and support the co-creation of consistent UML design models.

In practice, \texttt{Harmony Validator} improves the daily work of both educators and developers. 
In academic environments, it assists instructors in demonstrating the principles of consistency and integrity in UML design, allowing students to immediately visualize inconsistencies and learn corrective modeling behaviors through practice. 
In industrial settings, the tool enhances modeling workflows by automating consistency checks, reducing manual review effort, and increasing the overall reliability of model-driven development pipelines.

\section{Conclusions and Future Work}
\label{sec:conclusions}

This paper presented the \texttt{Harmony Validator}, an event-driven tool integrated into the Papyrus environment to detect and manage inconsistencies in UML class and sequence diagrams through real-time feedback and automated analysis. Its distributed, microservice-based architecture ensures scalability, modularity, and efficient event processing, while bridging modeling and verification to maintain consistent, reliable UML artifacts. Beyond its technical contribution, \texttt{Harmony Validator} advances education, research, and industrial practice: it helps students understand and correct modeling inconsistencies, enables researchers to collect empirical data on modeling behavior, and improves productivity and model integrity in professional workflows. By combining modern architectural patterns with intelligent feedback, the tool enhances the quality, traceability, and maintainability of UML models. Future extensions will broaden diagram support, refine detection rules, and incorporate learning-based mechanisms, reinforcing \texttt{Harmony Validator}’s role as a foundation for intelligent, context-aware modeling environments that foster human–AI collaboration in software design.

Future developments of the \texttt{Harmony Validator} will focus on extending its applicability, intelligence, and integration capabilities. 
Planned enhancements include adding new consistency rules and supporting a broader range of UML diagrams, such as activity, component, and state machine diagrams, to enable more comprehensive validation coverage. 
The tool will also incorporate \textit{machine learning} and \textit{large language model} components to suggest corrective actions, anticipate modeling errors, and automatically generate textual UML representations, reinforcing its synergy with tools such as \textit{PlainUML}. 
From a research perspective, \texttt{Harmony Validator} will continue to serve as a platform for empirical studies that analyze user behavior, cognitive processes, and learning patterns during modeling tasks. 
Furthermore, its containerized architecture will facilitate integration into continuous development environments, allowing future collaborations with industry partners to deploy the tool at scale in real-world software modeling pipelines.

\section{Acknowledgment}

This work was partially supported by the Conselho Nacional de Desenvolvimento Cient\'{i}fico e Tecnol\'{o}gico (CNPq) under Grant 312320/2025-6.

\bibliographystyle{ACM-Reference-Format}
\bibliography{sample-base}

\end{document}